\documentclass[twocolumn,showpacs,preprintnumbers,amsmath,amssymb]{revtex4}
\usepackage{graphicx}
\usepackage{dcolumn}% Align table columns on decimal point
\usepackage{latexsym}
\usepackage{dsfont}
\usepackage{bm}% bold math
\begin{document} 
\title{Fermion loop simulation of the lattice Gross-Neveu model}
\author{Christof Gattringer$^a$} 
\author{Verena Hermann$^{a,b}$}
\author{Markus Limmer$^a$}
\affiliation{
\vspace{3mm}$^a$Institut f\"ur Physik, FB Theoretische Physik,
Universit\"at Graz
\vskip0mm 8010 Graz, Austria
\vskip1mm$^b$Department of Earth and Environmental Sciences, Geophysics,
Munich University
\vskip0mm
 80333 Munich, Germany}
%%%%%\date{\vspace{2mm}February 24, 2007}

\begin{abstract}
%\vspace{4mm}
We present a numerical simulation of the Gross-Neveu model on 
the lattice using a new representation in terms of fermion loops. 
In the loop representation all signs due to Pauli statistics are 
eliminated completely and the partition function is a sum
over closed loops with only positive weights. We demonstrate that 
the new formulation allows to simulate volumes which are 
two orders of magnitude larger than those accessible with standard
methods. 
\end{abstract}

\pacs{11.15.Ha, 11.10.Kk}
\keywords{Pauli statistics, fermion loops, Gross-Neveu model}
\maketitle

\section{Introduction}
Numerical simulations with fermions are notoriously difficult.
The reason is that the minus signs due to Pauli statistics give rise
to cancellation effects. In 
quantum field theories the fermions are usually integrated out 
and the fermion determinant appears as a weight factor. Even in 
cases where the fermion determinant is real and positive its 
numerical treatment is very costly, since it 
essentially couples all degrees of freedom with each other   
and the individual contributions have changing signs.  

Finding alternative strategies for dealing with fermions would 
considerably improve the quality of numerical simulations.
Such strategies could either be new algorithms 
(see, e.g., \cite{wiese} for a prominent example) or a 
reformulation of the problem. 
Here we discuss the latter: In  
\cite{gattringer1} an alternative 
representation was given for a two-dimensional 
fermionic quantum field theory, the Gross-Neveu model.
The partition function was 
rewritten as a sum over closed loops where each contribution
has a real positive weight. This allows for a 
new approach to simulate the model which avoids dealing 
with the fermion determinant. 

This paper presents the first test of the loop approach for the
Gross-Neveu model in a 
numerical simulation, and we explore the prospects and limitations
of using loop-type representations in a numerical simulation 
of a fermionic system. Our results demonstrate 
that the method is promising and it is worthwhile to pursue it
in higher dimensions. In higher dimensions loop representations
of quantum field theories are known, but so far have exclusively
been used in the strong coupling limit \cite{loopqcd}. Our study
here, although 2-dimensional, is performed at arbitrary coupling. 
We are currently exploring the generalization to higher dimensions
and believe that for certain four-fermi interactions,
representations similar to the one used here can be 
found and successfully applied in numerical simulations. 

For gauge theories the situation is more complicated. 
Since gauge fields are oriented one has to use
oriented loops dressed with the gauge links and complex phases 
appear.
This was seen in the Schwinger model, where a loop representation 
exists \cite{gattringer2}, but a numerical simulation  
suffers from the fermion sign problem. Upon 
going to the strong coupling limit, the sign problem disappears
\cite{salmhofer} and a numerical simulation with loops
again unleashes its power \cite{galasa}. 
Four-fermi interactions may be generated 
with a Hubbard-Stratonovich transformation
using scalar fields. These do not introduce 
complex phases and a loop representation without signs is possible.

\section{The loop representation}

We begin with discussing the Gross-Neveu model \cite{gnmodel}
and its loop representation. In the continuum the action of the Gross-Neveu 
model is given by $S =  S_F + S_S$ with 
\begin{eqnarray}
S_F & = & \int d^2 x \, \overline{\psi}(x) 
\big[ \gamma_\mu \partial_\mu + \varphi(x)
+ m \big] \, \psi(x) \; ,
\nonumber \\
S_S & = & \frac{1}{2g} \int d^2 x \, \varphi(x)^2 \; .
\end{eqnarray}
Here $\psi$ and $\overline{\psi}$ are Grassmann valued vectors of $N$ 
flavors of 2-spinors and we use vector/matrix notation for both, spinor and 
flavor indices. The Euclidean partition function is defined by integrating
over all fields,
\begin{equation}
Z \; = \; \int \prod_x d \varphi(x) \, d \overline{\psi}(x)
\, \psi(x) \, \exp( \, - S[\varphi,\overline{\psi}, \psi ] \, ) \; .
\label{partitionfunction}
\end{equation} 
Upon integrating out the scalar fields $\varphi$, the model turns into a purely
fermionic theory with a four fermi interaction given by
$- g/2 \int d^2x ( \overline{\psi}(x) \psi(x) )^2$. The Gross-Neveu model 
is well understood analytically (see e.g.\ \cite{erlangen}) and has been
analyzed on the lattice in various settings \cite{latticegn}.

As it stands, the path integral (\ref{partitionfunction}) is only formally 
defined and a cutoff needs to be introduced. Here we use lattice 
regularization, which replaces the Euclidean space time $\mathds{R}^2$
by a finite regular lattice $\Lambda$. The path integral 
(\ref{partitionfunction}) is well defined when the measure is understood
as the product over individual measures over the fields living
on the lattice points. The action is discretized using the Wilson
formulation such that it reads
\begin{eqnarray}
S_F & = & \sum_{x \in \Lambda} \overline{\psi}(x) 
\bigg( -\! \sum_{\mu = \pm 1}^{\pm2} 
\frac{1 \mp \gamma_\mu}{2} \, \psi(x \pm \hat{\mu}) 
\nonumber \\
& & \qquad \qquad
+ \; \varphi(x) \, \psi(x) \; + \; [ 2 + m ] \, \psi(x) \bigg) \; ,
\nonumber \\
S_S & = & \frac{1}{2g} \sum_{x \in \Lambda} \varphi(x)^2 \; .
\label{latticeaction}
\end{eqnarray} 
For the scalar field $\varphi$ we use periodic boundary conditions 
for both directions, the fermions are periodic in the spatial direction
and anti-periodic in time. 

Using hopping expansion techniques, the $N$-flavor lattice 
Gross-Neveu model (\ref{latticeaction}) can be mapped 
into a model of 2$N$ sets of loops \cite{gattringer1}. 
For convenience we will often refer to the loops in different sets 
as blue, red etc.\ loops. Within each set the loops are non-oriented, 
closed and self-avoiding. However, when loops belong to different sets,
e.g., a red and a blue loop, they may touch or cross each other. 
The partition function of the lattice Gross-Neveu
model is then a sum over all possible configurations of the loops
in the 2$N$ sets. Each configuration has a positive weight 
computed from the loops. 

Although \cite{gattringer1} gives the 
partition function for arbitrary $N$, we here only quote the one-flavor
expression which we use in our simulation. For $N = 1$ we need two sets 
of self-avoiding loops, red and blue, denoted by $r$ and $b$. 
The one-flavor partition function in the loop representation reads
(up to an overall normalization factor)
\begin{equation}
Z \; = \; \sum_{r,b} \, \left(1/\sqrt{2}\,\right)^{c(r,b)} \, 
f_1^{n_1(r,b)} \, f_2^{n_2(r,b)} \; .
\label{zloop}
\end{equation}
In this formula $c(r,b)$ is the total number of corners for both
red and blue loops. Thus every corner contributes a factor of 
$1/\sqrt{2}$ to the weight of a configuration.
$n_1(r,b)$ is the number of lattice sites which 
are singly occupied by either $r$ or $b$ and 
$n_2(r,b)$ is the number of doubly occupied sites. 
We remark that, since the loops in the two sets are self avoiding, 
double occupation can appear only when a red and a blue loop 
cross our run alongside each other. The weight factors $f_1$ and $f_2$
are related to the mass $m$ and the coupling $g$ through
\begin{equation}
f_1 \, = \, (2+m)[(2+m)^2 + g]^{-1} \; , \; f_2 \; = \; [(2+m)^2 + g]^{-1} \; .
\label{f1f2}
\end{equation}

We stress that the mapping (\ref{zloop}), (\ref{f1f2}) 
is exact in the thermodynamic
limit. For finite volume different types of boundary conditions in the two 
representations lead to finite size effects: In the loop 
representation we need to have closed loops and in a finite volume
the loops can wind around the compact lattice. The loop configurations
fall into three equivalence classes, $C^{ee}, C^{eo}, C^{oo}$, depending on 
the numbers of red and blue non-trivially winding loops  
(see also \cite{galasa}): 
$C^{ee}$ (even-even): The total number of windings for both, 
red and blue loops is even for both directions. $C^{eo}$ (even-odd): 
One of the colors has an 
odd number of windings for one of the directions. $C^{oo}$ (odd-odd): 
Both colors
have an odd number of windings in one of the directions. These 
equivalence classes cannot be linked in a simple way to the boundary 
conditions in the standard representation which we discussed above.
However, below we will demonstrate that the boundary effects vanish as
$1/\sqrt{V}$, with $V$ denoting the volume. 

\section{Numerical simulation}

The numerical simulation of the Gross-Neveu model now is performed
directly in the loop representation (\ref{zloop}) using a local Metropolis
update (see e.g.\ \cite{landaubinder}). We update the red and the
blue loops alternately, by performing a full sweep through the lattice for
one of the colors and treating the other one as a background field. 
A sweep consists of visiting all plaquettes of the 
lattice. For each plaquette we generate a trial configuration by
inverting the occupation of the color we want to update 
for all the links in the plaquette. This guarantees that all loops remain
closed. Furthermore new loops may be generated when all links of the plaquette
are empty. When the trial configuration violates the self-avoiding
condition it is rejected immediately and the algorithm tries the next 
plaquette. Otherwise the trial configuration is accepted with the 
Metropolis probability 
$p = ( 1/\sqrt{2} )^{\Delta c}  
f_1^{\Delta n_1} f_2^{\Delta n_2}$,
where $\Delta c$ is the change in the number of corners and 
$\Delta n_1, \Delta n_2$ are the changes in the occupation numbers.
The initial configuration can either be the empty lattice (for $C^{ee}$) 
or has one or two winding loops ($C^{eo}$ and $C^{oo}$).

The observables we discuss here are all first and second derivatives 
of the free energy $F = -\ln Z$, and can be written 
as moments of the occupation numbers. In particular
for the chiral condensate and its susceptibility, which in the standard 
language are given by 
\begin{equation}
\chi \; = \;  
\frac{1}{V} \sum_{x \in \Lambda} \langle \, 
\overline{\psi}(x) \psi(x) \, \rangle \; = \;
- \frac{1}{V} \frac{ \partial \ln Z}{\partial m} \; \; \; , \; \; \;
C_\chi \; = \; \frac{\partial \chi}{\partial m} \; ,
\end{equation}
we quote the corresponding expressions in terms of occupation numbers
and their fluctuations,
\begin{eqnarray}
\chi & = & - \, \frac{1}{V \, f_1} \, \Big( f_2 \langle n_1 \rangle
\, + \, 2 f_1^2 \langle n_0 \rangle \Big) \; ,
\nonumber \\
C_\chi & = &  - \, \frac{1}{V \, f_1^2} \, \Big( [4f_1^4 - 2f_1^2 f_2]\, 
\Big\langle (n_0 - \langle n_0 \rangle )^2 \Big\rangle 
\nonumber \\
& & + \; [f_2^2 - 2 f_1^2 f_2] \,  
\Big\langle (n_1 - \langle n_1 \rangle )^2 \Big\rangle 
\nonumber \\
& & + \; 2 f_1^2f_2 \, 
\Big \langle (n_0 + n_1 - \langle n_0 + n_1 \rangle)^2 \Big\rangle
\nonumber \\
& & 
- \; [4f_1^4 - 2f_1^2 f_2] \langle n_0 \rangle - f_2^2 \,
\langle n_1 \rangle \Big) \; .
\label{occobs}
\end{eqnarray}
Here we have introduced $n_0$, the total number of empty
sites, i.e, sites visited by neither a red nor a blue loop. 

Equivalent formulas can be derived for the internal energy, 
the heat capacity as well as for derivatives of the free energy
with respect to the coupling $g$.
$n$-point functions may be treated as usually by introducing 
source fields and differentiating with respect to them.
This gives expressions involving correlators of {\sl local}
occupation numbers. Finally, the generalization of the 
above formulas to an arbitrary number of flavors is straightforward.

\section{Results}

\begin{figure}[t]
\begin{center}
\includegraphics[width=85mm,clip]{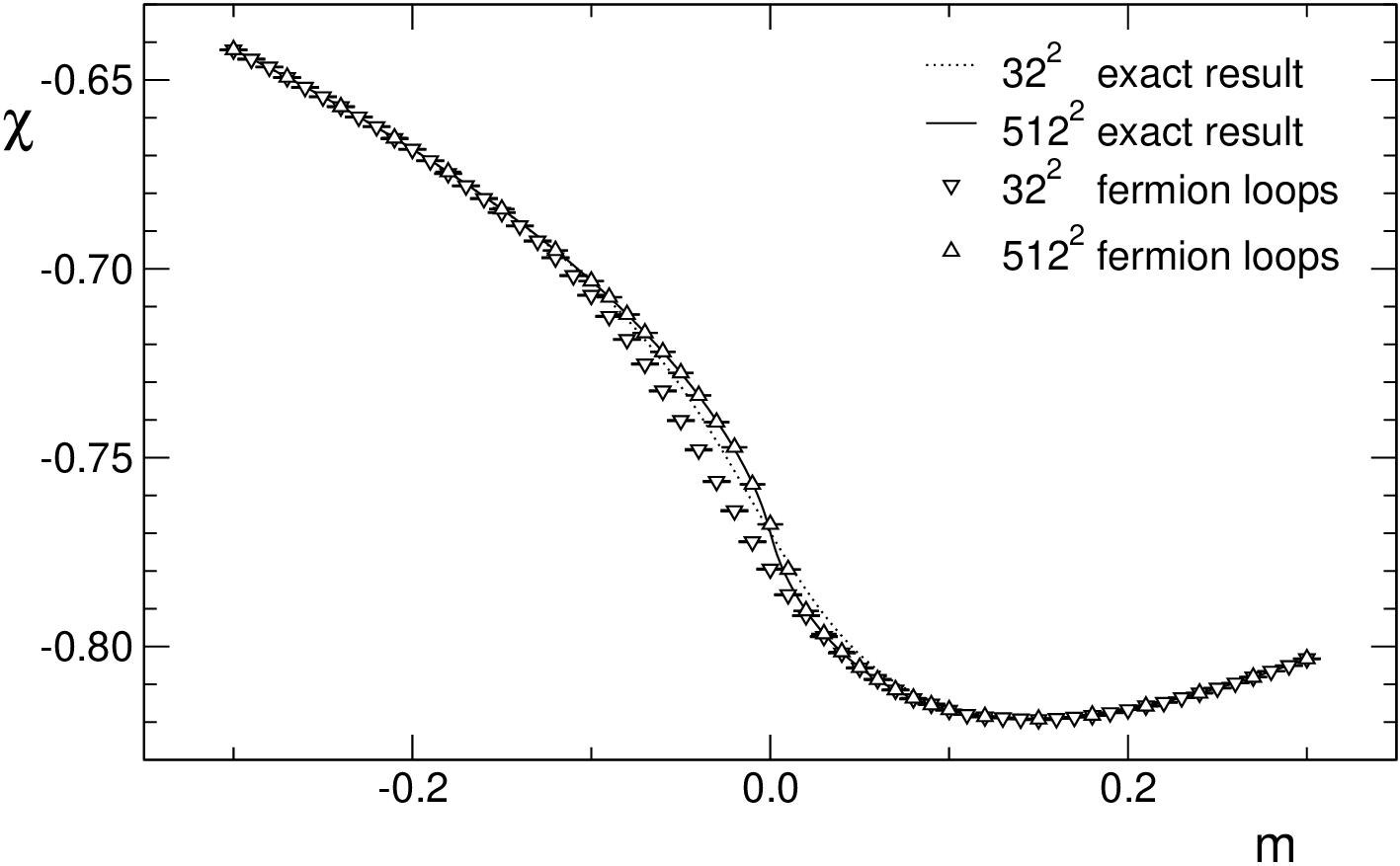} 
\end{center}
\caption{The chiral condensate $\chi$ for $g=0$ as a function of $m$
for 2 different lattice sizes. We compare the simulation 
in the loop representation (symbols with error bars) to the exact 
result from Fourier transformation (curves).}
\label{fig1}
\end{figure}

In this section we present some selected results which serve to 
illustrate the advantages of the loop approach, but also allow to assess 
its limitations. In order to compare with traditional methods, we 
performed a reference simulation of the Gross-Neveu model using standard
methods. The fermions were integrated out giving rise to the fermion
determinant in a background configuration of the scalar field $\varphi$. 
These background configurations were computed according to the Gaussian 
distribution of $S_S$, and the determinant was used as a factor for
reweighting. This is possible, since the eigenvalues of the Dirac matrix
come in complex conjugate pairs, and the scalar field does not have
topological modes which could give rise to zero eigenvalues. Thus the
fermion determinant is always strictly positive.

We stress that the reweighting in the standard formulation
works only in two dimensions
due to the numerical cost of evaluating the determinant. However, for our 
problem where the scalar fields are independent Gaussians at each site, 
reweighting has the big advantage, that autocorrelation is avoided. 
Alternative strategies such as Hybrid Monte Carlo, cannot make use of that
advantage. 

Another important conceptual point has to be addressed: For the free case,
$g = 0$, the standard representation allows for an exact solution with the help
of Fourier transformation. In the loop formulation, 
however, the case $g=0$ is not
special at all. Thus $g=0$ is the optimal point for testing the power of 
the loop approach to the limits because we have exact results on 
almost arbitrary large volumes, which we use to compare
with the data of the loop simulation.
Since the weight factors $f_1$ and $f_2$ of Eq.\ (\ref{f1f2})
are smooth functions of $g$ and $m$ it is reasonable 
to transfer the experience obtained with 
the loop approach at $g=0$ to nearby values of $g$.

\begin{figure}[t]
\begin{center}
\includegraphics[width=85mm,clip]{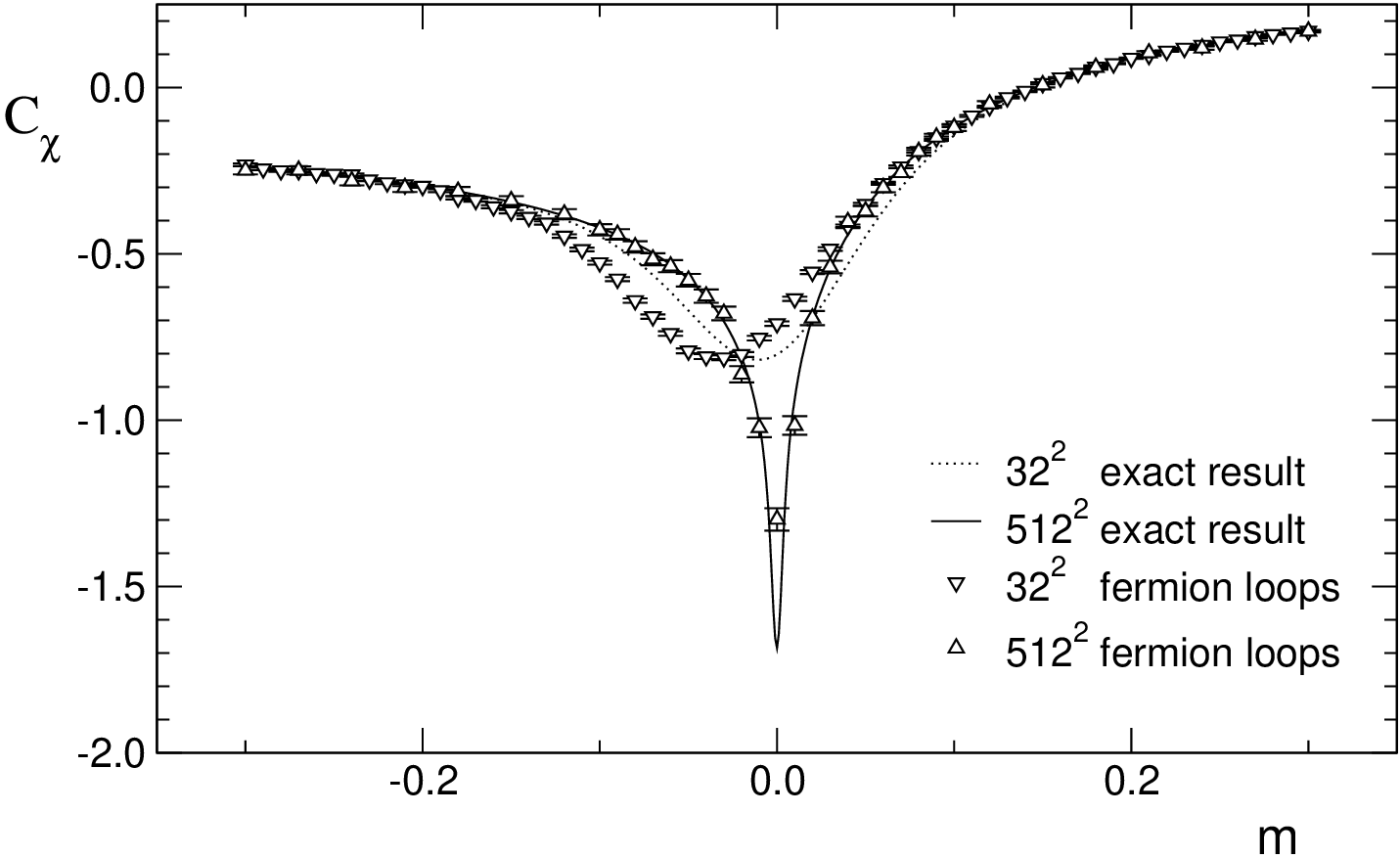} 
\end{center}
\caption{Same as Fig.\ 1, now for the chiral susceptibility $C_\chi$.}
\label{fig2}
\end{figure}

Thus we begin our assessment of the loop approach at $g=0$. In Figs.\ 1 and 2
we compare the loop results in the $C^{ee}$ sector (symbols) with 
those from Fourier transformation 
(curves). We use two volumes for the comparison, a relatively small lattice 
of size $32 \times 32$ and a considerably larger one, $512 \times 512$.
For the simulation in the loop approach at each value of $m$ 
we typically performed $10000$ sweeps of our local update for both colors for 
equilibration and used $50000$ measurements of the observables separated 
by $10$ pairs of sweeps. The observables were calculated using the 
occupation number representation (\ref{occobs}) and the statistical 
error was computed with the jackknife method.

Fig.\ 1 shows that already on the small lattice the data points 
are very close to the exact result. The largest discrepancy is seen near
$m=0$, the chiral point where the fermions become massless. For the larger
lattice the data points fall exactly on top of the analytic result.
The situation is similar for the susceptibility in Fig.\ 2.
For the small lattice we find a clear finite size effect, a shift of the 
susceptibility curve. On the larger lattice the agreement is almost perfect and 
only at the chiral point $m=0$ we still see a 
slight discrepancy. We remark, that the decrease of the minimum of $C_\chi$ 
with increasing volume $V$ does not signal a phase transition (at $N=1$ 
there is no discrete symmetry that could be broken spontaneously).
The minima decrease only logarithmically with $V$. For $g=0$ it can be shown 
exactly that $C_\chi$ diverges logarithmically when removing the IR cutoff.
For $g > 0$ we could fit the minimum of $C_\chi$ as
obtained from the simulation very reliably with $\ln V$.  
Also the comparison with the results from the standard
approach shows that the largest discrepancy is found near the chiral point,
which, however, vanishes quickly with
increasing volume.

An important part of comparing the standard and the loop
approach is to test how the different types of boundary effects 
scale with the volume and at what rate the perfect equivalence of the 
two representations is reached with increasing $V$. We assess this question 
directly in the loop approach: At a fixed point $(m,g)$ in parameter 
space we compute the chiral condensate $\chi$ for the three different 
equivalence classes $C^{ee}, C^{eo}, C^{oo}$ introduced above. 
This is repeated 
for several volumes $V$ and in Fig.\ 3 we plot the discrepancy of the results
as a function of $\sqrt{V}$. The data shown in the plot are for
$g = 0.1$ and $m = 0.0$.
In the plot $\Delta \chi^{(eo-ee)}$ denotes the
discrepancy between the $C^{eo}$ and $C^{ee}$ results and 
$\Delta \chi^{(oo-ee)}$ is the splitting between $C^{oo}$ and $C^{ee}$.
Also a comparison of $\chi$ to the results from the standard approach with
mixed boundary conditions shows a 1/$\sqrt{V}$ behavior.

\begin{figure}[t]
\begin{center}
\includegraphics[width=85mm,clip]{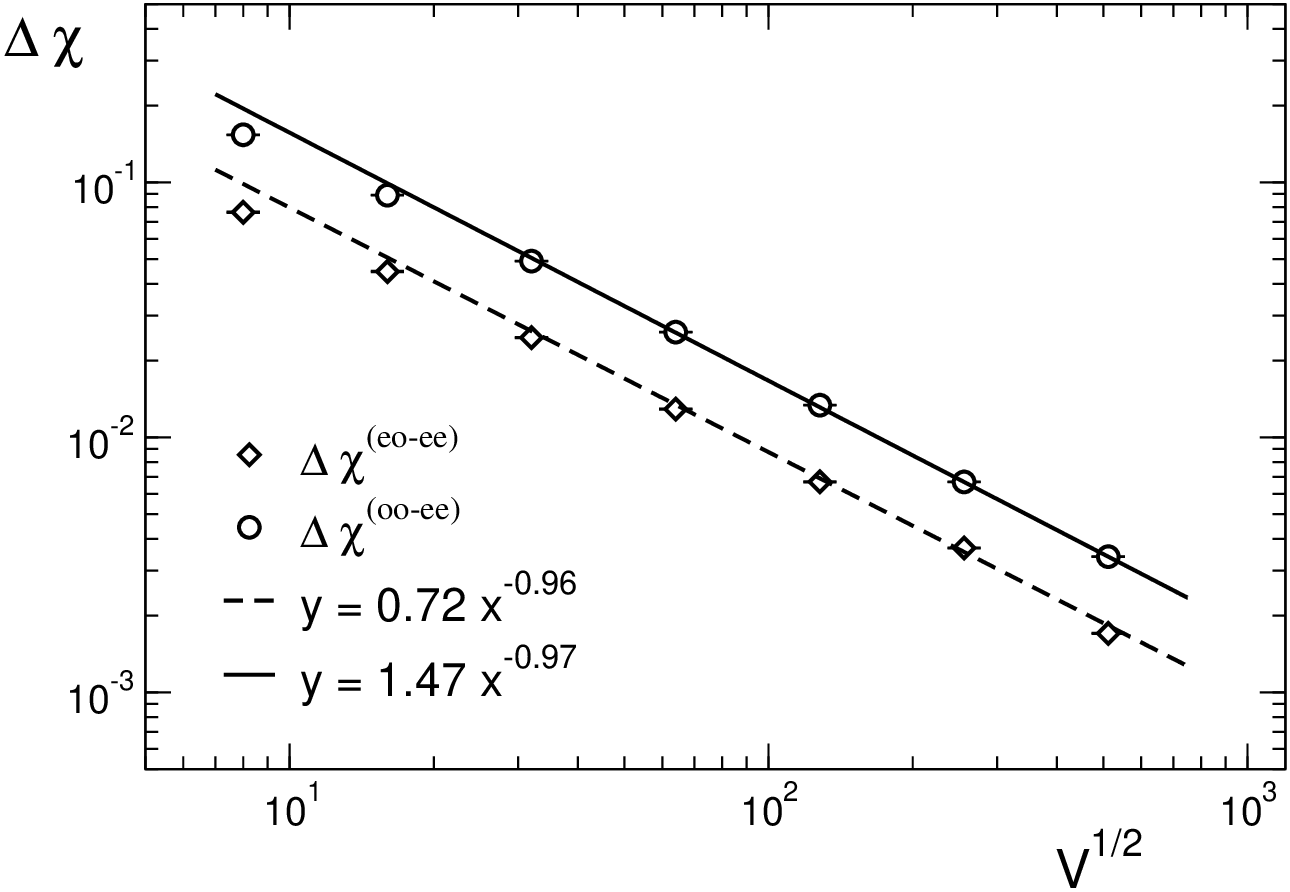} 
\end{center}
\caption{
Splitting of the results for the chiral condensate in the 
different equivalence classes as a function of the volume.}
\end{figure}

The splitting $\Delta \chi$ can be analyzed with the help of mean field
theory. One finds that it should behave as $\Delta \chi \propto 1/\sqrt{V}$. 
We performed a fit of our data to the form $\Delta \chi \propto c V^{\alpha}$
and found values of $\alpha$ which are close to $-1/2$ for all analyzed values of 
$(m,g)$ (see also Fig.\ 3). 
Thus mean field arguments as well as our numerical 
findings indicate, that the finite volume effects scale as $1/\sqrt{V}$.

\vfill

\section{Discussion}

In this letter we have explored an alternative formulation for fermionic 
systems using the example of a 
2-dimensional quantum field theory. The representation in
terms of fermion loops allows one to simulate the system without having to use
fermion determinants. An important aspect is that in the loop
formulation used here
we are not restricted to the case of strong coupling but can work at arbitrary
$g$. In this exploratory study we simulate the model using 
a simple local update and compare the outcome to analytic results and 
the data from a simulation in the standard approach. Many observables can be
expressed in terms of occupation numbers and their correlators. 
We show that finite size effects decrease like
$1/\sqrt{V}$ and thus the thermodynamic limit, where the loop representation
becomes exact is approached rapidly. 

An important issue is of course the assessment of the gain in numerical
efficiency when using the loop algorithm. Already with the local algorithm
used here a considerable increase of the accessible volumes was found. 
Using the same small cluster of PC's the standard approach could be used on
lattices with a maximum volume of $32 \times 64$, while in the loop
formulation we were able to simulate systems up to $700 \times 700$, which 
is an increase of the volume by more than two orders of magnitude. This
enormous improvement is a strong incentive to search for loop representations 
also in higher dimensional fermion systems and for four-fermi interactions
no principal obstacles seem to appear.

{\bf Acknowledgments:} 
We thank Erek Bilgici, Philipp Huber, Christian Lang, Klaus Richter, Andreas
Sch\"afer and Erhard Seiler for discussions and helpful remarks.

\clearpage
\end{document}